\documentclass[10pt,twoside,british]{ilcws08}
\usepackage[latin1]{inputenc}
\usepackage[dvips]{graphicx,epsfig,color}
\usepackage{wrapfig,rotating}
\usepackage{amssymb,amsmath,array}

\begin{document}

\title{Sensitivity to the Higgs Self-coupling Using the ZHH Channel}

\author{Michele Faucci Giannelli\vspace{0.3cm}\\
Royal Holloway University of London}

\maketitle
\begin{abstract}
The Standard Model predicts the value of the Higgs self-coupling but
it cannot be measured at LHC. This measurement requires a machine
such as the proposed International Linear Collider. Here, the sensitivity
to the Higgs self-coupling is evaluated using the ZHH to six jets
channel for a Higgs mass of 120\,GeV/c$^{2}$. Full simulation has
been carried out for an integrated luminosity of 500\,fb$^{-1}$.
Several analyses are presented and all evaluate the cross section
resolution to be about 180$\%$. Potential areas for improvement are
identified.
\end{abstract}

\section{Introduction}

At the energy of the International Linear Collider (ILC), the process
$\mathrm{e^{+}e^{-}\rightarrow ZHH}$ is the only one that can be
used to measure the self-coupling of the Higgs boson. This study was
performed assuming $\mathrm{M_{H}}=120\,$GeV and a centre-of-mass
energy of 500\,GeV at which the total cross section is 0.183\,fb,
assuming a polarisation of -80\% for the electron beam. The main decay
mode is the six-jet final state with a BR of about $40\%$, which
is the only final state considered in this analysis. The integrated
luminosity was assumed to be 500\,fb$^{-1}$ which corresponds to
the planned integrated luminosity for the first phase of ILC.

\section{Generation, simulation and reconstruction}

The events used in the ZHH analysis were generated using Pandora Pythia
\cite{Pandora} and WHIZARD \cite{WHIZARD}. The two generators were
compared and are compatible. Since it is computationally impossible
to perform the simulation for the whole 500\,fb$^{-1}$, only events
with six quarks in the final state and a selection of four-jet final
states were considered. Table \ref{tab:List-of-events}%
\begin{table}
\begin{tabular}{|c|c|c|c|c|c|c|c|c|c|c|c|}
\hline 
Channel&
ZHH&
$\mathrm{t\bar{t}}$&
WWZ&
ZZH&
ZZZ&
ZZ&
ZH&
tbtb&
Wtb&
ttH&
ttZ\tabularnewline
\hline
\hline 
$\sigma$(fb)&
0.183&
711&
212.9&
0.502&
1.486&
90.5&
13.66&
0.434&
44.34&
0.237&
1.016\tabularnewline
\hline 
Events &
10k&
375k&
120k&
1000&
5000&
50k&
20k&
5000&
25k&
5000&
5000\tabularnewline
\hline
\end{tabular}

\caption{\label{tab:List-of-events}Signal and principal background channels
and number of events generated for each.}
\end{table}
 summarises the events generated. The main background is the hadronic
$\mathrm{t\bar{t}}$ channel which has a total cross section of 326\,fb.
For the ZZ channel it was required to have at least one Z decaying
in heavy quarks (c, b). 

The detector simulation was performed using MOKKA v00-06-04p02. The
detector model used was LDC00Sc \cite{detector model}.

The simulated events were reconstructed using Marlin v00-09-10. The
hits in the tracking and calorimetry systems were digitised and then
used as input for the tracking (FullLDC package) and particle flow
reconstruction (PandoraPFA package). The particles were forced to
six jets using the Durham algorithm and the jets were analysed by
the vertex reconstruction software (LCFI package) to reconstruct b
and c vertices. The same reconstruction chain was performed using
the {}``perfect'' particle flow reconstruction, in which all particles
are correctly reconstructed, essentially neglecting any confusion
term from the calorimeters. Details of all the software used can be
found in \cite{Marlin Reco}. The reconstructed particles and the
jets, for both realistic and perfect PFA chain of reconstruction,
were then used to calculate several shape variables which were used
in this analysis

\section{Cut based analysis}

A key variable used to separate signal from background was the sum
of all outputs from the b tagging neural network. This is a number
from 0 to 1 for each jet, where 1 indicates b-like jets and 0 light-like
jets. Having 6 jets, the variable used in the analysis varies between
0 and 6 with the signal peaking at 4 and the main background at 2.
The other variables used were: thrust, cos$\theta_{thrust}$, second
Fox-Wolfram moment, total energy, number of tracks, number of particles
in jet, angular distance between jets (Y6) and jet EM energy ratio.
These variables were optimised by scanning simultaneously all variables
over a wide range of values. However the high number of variables
made it difficult to test many values because of the processing time
and the memory requirement. A satisfactory compromise was found using
five cut values for each variable and reiterating the process to find
the exact maximum. For each iteration, among all possible combination
of cuts (there are $5^{8}$), the one that maximised the usual figure
of merit $\mathrm{S/\sqrt{S+B}}$ was chosen. After few iterations
the value of $\mathrm{S/\sqrt{S+B}}$ did not improve any further
and the process was ended. The final value for $\mathrm{S/\sqrt{S+B}}$
after applying all the cuts was $0.364\pm0.011$. A similar optimisation
was performed for the perfect PFA reconstruction obtaining $\mathrm{S/\sqrt{S+B}}=0.361\pm0.010$.

In order to further separate signal from background, a $\chi^{2}$
was built to force the reconstruction of each event to ZHH:

\begin{equation}
\chi^{2}=\frac{(M_{ij}-M_{Z})^{2}}{\sigma_{Z}^{2}}+\frac{(M_{kl}-M_{H})^{2}}{\sigma_{H}^{2}}+\frac{(M_{mn}-M_{H})^{2}}{\sigma_{H}^{2}}.\label{eq:chi}\end{equation}

All forty five combinations of the six jets were tried. The combination
that produced the smallest $\chi^{2}$ defined the $\chi_{min}^{2}$
for that event. $\chi_{min}^{2}$ was then used to discriminate signal
from backgrounds. However the large number of $\mathrm{t\bar{t}}$
passing the previous cuts can be reconstructed to look similar to
ZHH events due to the high combinatorial in jet pairing. For this
reason a second $\chi^{2}$ was built adding the b tagging information:

\[
\chi^{2}=\frac{(M_{ij}-M_{Z})^{2}}{\sigma_{Z}^{2}}+\frac{(M_{kl}-M_{H})^{2}}{\sigma_{H}^{2}}+\frac{(M_{mn}-M_{H})^{2}}{\sigma_{H}^{2}}+\sum_{J_{H}=1}^{4}A(Btag(J_{H})-1)^{2}.\]

The new term uses the b tag information with the value of $A$ which
has been found to be very large after an optimisation. Since the four
jets from the two Higgs bosons should be b jets, the output of the
b tagging neural network should peak at 1 hence the sum of the four
terms should peak at zero for well reconstructed and well associated
jets in signal events. Since b-like jets are more likely to form one
of the Higgs boson instead of the Z, the new term effectively reduces
the number of possible combinations. This reduction in combinations
has a small impact on the signal but reduces all the backgrounds. 

\begin{wrapfigure}{R}{0.5\columnwidth} 
\centering 
\includegraphics[scale=0.40]{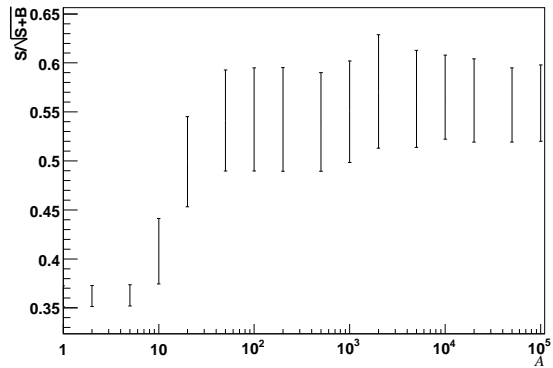}
\caption{$\mathrm{S/\sqrt{S+B}}$ as a function of $A$.} 
\label{fig:ConstB nofit} 
\end{wrapfigure}

The optimisation of the parameter $A$ was performed varying the value
from 0 to $10^{5}$. For each value of $A$ the same procedure described
before was performed; the minimised $\chi^{2}$ was plotted for signal
and background and from this distribution the $\mathrm{S/\sqrt{S+B}}$
was maximised. Figure \ref{fig:ConstB nofit}    shows the maximum of $\mathrm{S/\sqrt{S+B}}$ as a function of
$A$. The error is the statistical error mainly due to the limited
number $\mathrm{t\overline{t}}$ events. For values above 100 the
separation is constant at a value of $0.55\pm0.06$. 

Kinematic fit of the six jets to further constrain the reconstruction
was implemented but the separation was not improved. This and the
fact that the optimisation of $A$ is asymptotic and does not have
a peak, are indications that the mass resolution is less important
than the b tagging performance. A separate analysis on the events
with perfect PFA reconstruction confirmed this indication. In fact
the separation for the perfect reconstruction is $\mathrm{S/\sqrt{S+B}}=0.59\pm0.06$,
which, within the statistical error, is compatible with the case of
realistic PFA. This means that any improvement in the PFA will not
reflect in a better separation in this analysis. In order to have
a better separation the other main element of the selection, the vertex
reconstruction for the b tagging, has to be improved.

\section{Neural network analysis}

A second analysis was performed using a neural network, which in principle
should give a better separation between the signal and the backgrounds
then the cut based one. The network implementation was performed using
the artificial neural network (ANN) package within TMVA \cite{TMVA}.
A separate sample of signal and background events was generated to
train the neural network. For the background, an integrated luminosity
of $125\mathrm{\, fb^{-1}}$ was generated while for the signal 30000
events were used. The preliminary cuts described above were applied
to the training sample and the events passing the cuts were used to
train the network. Due to the limited number of events left, only
a simple network could be trained; in particular two configurations
were studied. The variables used were the b tagging, the $\chi_{ZHH}^{2}$
and the $\chi_{t\overline{t}}^{2}$. The two $\chi^{2}$ variable
are defined in Eq. \ref{eq:chi} but the b jets were forced to came
from the Higgs or from the decay of the top. 

The $\mathrm{S/\sqrt{S+B}}$ were obtained as before, looking for
the maximum as a function of the neural network output; the results
are summarised in Table \ref{tab:Best separation}. \begin{wraptable}{L}{0.75\columnwidth} 
\begin{tabular}{|c|c|c|c|} 
\hline 
Analysis &  $\mathrm{S/\sqrt{S+B}}$ & S & B \\ \hline 
Simple ${\chi^{2}}$ & ${0.36\pm0.01}$ & 13.5 & 1364.5 \\ \hline 
${\chi^{2}}$ with b tag term & ${0.55\pm0.06}$ & 4.0 & 47.0 \\ \hline 
${\chi^{2}}$ with b tag term and kin. fit. & ${0.56\pm0.06}$ & 6.4 & 124.4 \\ \hline 
NN two variables & ${0.57\pm0.06}$ & 5.8 & 99.2 \\ \hline 
NN three variables & ${0.55\pm0.06}$ & 7.5 & 186.0 \\ \hline 
\end{tabular}
\caption{\label{tab:Best separation} Best $\mathrm{S/\sqrt{S+B}}$ for different NN and cut based analyses.} 
\end{wraptable} Within the statistical error, neither of the two networks improved
the separation between signal and background. This is a further confirmation
that, at the moment, the mass information does not have any discriminating
power.

\section{Conclusion}

Given the relevance of the b tagging performance in the analysis,
a dedicated study was performed to study the performance in the six-jet
environment. The efficiency in b tagging for b, c and light jets was
compared between the two-jet and the six-jet environment. The comparison
showed an increase of about 25$\%$ in the fake rate from c jets while
the light jets in the six-jet environment had a fake rate doubled
with respect to the two-jet environment. This increase is due to the
different environment but also to the different energy of the jets.
A separate study of two-jet events of different energy showed that
the fake rate increases at higher energies with respect to the nominal
value obtained with jets from the decay of a Z boson at rest.

If a similar performance could be achieved in the six-jet environment
as in the two-jet events, the resolution would improve by a factor
two. Then, without performing any further optimisation, the resolution
on the ZHH cross section would be about 95\%. This value is not too
distant from those obtained in fast Monte Carlo studies, about 80\%
for \cite{T Barklow} and 60\% for \cite{P Gay}, if the same integrated
luminosity is considered and if effects such as the gluon emission
are considered. The remaining difference is likely to be due to detector
effects, such as confusion in particle reconstruction. It is important
to stress the fact that this is an indirect comparison; in order to
have an accurate estimate of the differences between fast and full
simulation, the same events should be compared using the same analysis.
Further improvement could be achieved considering also the decay of
the Z to neutrinos.

\end{document}